\documentclass[12pt]{article}

\usepackage{amsmath}
\usepackage{amssymb}
\usepackage{amsfonts}
\usepackage{graphicx}
\begin{document}

\begin{center}
{\large \bf{ The Ellipsoidal Universe and the Hubble tension }}
\end{center}

\vspace*{1.0 cm}

\begin{center}
{
Paolo Cea~\protect\footnote{Electronic address:
{\tt paolo.cea@ba.infn.it}}  \\[0.5cm]
{\em INFN - Sezione di Bari, Via Amendola 173 - 70126 Bari,
Italy} }
\end{center}

\vspace*{1.0 cm}

\begin{abstract}
\noindent 
The Hubble tension resides in a statistically significant discrepancy between early time and late time determinations
of the Hubble constant. We discuss the Hubble tension within the Ellipsoidal Universe cosmological model. We
suggest that allowing small anisotropies in the large-scale spatial geometry could alleviate the tension.
We, also, show that the discrepancy in the measurements of the Hubble constant is reduced to a statistically acceptable
level if we assume sizeable cosmological anisotropies during the Dark Age. In addition, we argue that the Ellipsoidal
Universe cosmological model should resolve the $S_8$ tension.
\end{abstract}

\vspace*{0.6cm}

\vspace{0.2cm}
\newpage
\noindent
\section{Introduction}
\label{s-1}
One of the most fundamental cosmological parameter is the Hubble constant $H_0$~\cite{Jackson:2007,Freedman:2010} that measures the 
current expansion rate of the Universe. Recently, a statistically significant discrepancy has emerged between different methods of measuring 
the Hubble constant.
Indeed, there is an evident tension between the Hubble parameter measured by late universe observations and the one measured by
the Plank Collaboration ( a fair exhaustive account can be found in the reviews Refs.~\cite{Knox:2020,Shah:2021,DiValentino:2021,Abdalla:2022} 
and references therein; see also Refs.~\cite{Dainotti:2021,Bargiacchi:2022,Dainotti:2022,Lenart:2023}). \\
\noindent
The Planck Cosmic Microwave Background (CMB) angular power spectra provided the most precise determination of the cosmological
parameters. To extract $H_0$ from the CMB data it is necessary to assume a model for the expansion history of the Universe. Actually,
the CMB measurements by the {\it Planck}  satellite have confirmed to a high level of accuracy the standard  $\Lambda$ Cold Dark Matter
($\Lambda$CDM) cosmological model based on the flat Friedmann-Robertson-Walker metric:
\begin{equation}
\label{1}
d s^2 \; = \; - c^2 \; dt^2 \; + \;  a^2(t) \; \delta_{ij} \;  dx^i  \; dx^j \; 
\end{equation}
with a cold dark matter component and a dark energy component in the form of a cosmological constant.
The 'TT, TE, EE + low E + lensing + BAO' best fit $\Lambda$CDM model to the Planck 2018 data~\cite{Aghanim:2021} 
furnished for the Hubble constant:
\begin{equation}
\label{2}
H_0  \; =  \;  66.76 \; \pm \; 0.42 \; \; km \; s^{-1} \; Mpc^{-1} \; 
\end{equation}
at the 68 \% confidence level. On the other hand, the Supernovae H0 for the Equation of State (SH0ES) Team reported the most recent local
measurement of H$_0$ obtained by the cosmic ladder of Cepheid-SN Ia standard candles~\cite{Riess:2021}:
\begin{equation}
\label{3}
H_0  \; =  \;  73.04 \; \pm \; 1.04 \; \; km \; s^{-1} \; Mpc^{-1} \;  \; \; .
\end{equation}
These two independent estimates of the Hubble constant are in tension with each other at a significant statistical level reaching
about five standard deviations.  However, it should be mentioned that the debate over the value of the 
Hubble constant is  not yet settled and we must wait for future measurements 
to finally reach to a resolution of the current discrepancies. Indeed,
there has  been mounting evidence from local calibrations through 
the tip of the red giant branch (TRGB) that the Hubble tension is either
non-existent or at least much smaller than the discrepancies between SH0ES and the 
Planck determinations~\cite{Freedman:2019,Freedman:2020,Freedman:2021}. \\
In addition to the already mentioned Hubble constant disagreements, a tension between the Planck data with weak leasing measurements and 
the redshift surveys has been reported about the value of the parameter $S_8 = \sigma_8 \sqrt{\frac{\Omega_m}{0.3}}$, where
$\Omega_m$ is the present time value of the nonrelativistic matter density and $\sigma_8$ is the amplitude of growth of structures.
As a matter of fact, it is now well established that this $S_8$ tension is driven by $\sigma_8$ rather than $\Omega_m$.
To be concrete, here we report the recent measurement from a joint cosmological analysis of weak gravitational lensing observations
from the Kilo-Degree Survey (KiDS-1000), with redshift-space galaxy clustering observations from the baryon Oscillation Spectroscopy Survey
(BOSS) and galaxy-galaxy lensing observations. The combination between KiDS-1000, BOSS and the Spectroscopic 2-degree Field
Lensing Survey (2dFLenS), presented in Ref.~\cite{Heymans:2021}, resulted in the following constraint on the structure growth parameter:
\begin{equation}
\label{4}
S_8 \;   =  \;   0.766~^{+ 0.020}_{- 0.014}  \; \; .
\end{equation}
This value of the $S_8$ parameter should be compared with the one estimated by the Planck Collaboration within the standard $\Lambda$CDM
cosmological model~\cite{Aghanim:2021}:
\begin{equation}
\label{5}
 S_8 \;   =  \;   0.825 \; \pm \; 0.011   \; \; .
\end{equation}
From Eq.~(\ref{5}) we see that there is a mismatch of about three standard deviations between the $S_8$ value estimated by the
Planck Collaboration and the value reported in Eq.~({\ref{4}). 
\\
In the present note we perform an exploratory study of these  cosmological tensions within the Ellipsoidal Universe cosmological 
model. \\
The remaining part of the paper is organised as follows. In Sect.~\ref{s-2}
we  briefly present the Ellipsoidal Universe cosmological model.  Sect.~\ref{s-3} is devoted to The Hubble tension within the 
Ellipsoidal Universe model, while  the $S_8$ tension is discussed in Sect.~\ref{s-4}.
Finally, in Sect.~\ref{s-5} we, briefly, summarise the results presented in this paper and draw our conclusions.
\section{The Ellipsoidal Universe}
\label{s-2}
The ellipsoidal Universe cosmological model~\cite{Campanelli:2006,Campanelli:2007} was proposed to cope with several anomalous 
features at large scales in the cosmic microwave background anisotropy data. 
Indeed, even the Planck 2018 data confirmed the presence of large-scale anomalous features.
As it is well known, the most evident anomaly concerned the quadrupole temperature correlation that was suppressed with respect to
the best-fit $\Lambda$CDM cosmological model. In Refs.~\cite{Campanelli:2006,Campanelli:2007} it was suggested that, if one allows
the large-scale spatial geometry of the Universe to be only plane-symmetric, then the quadrupole amplitude can be drastically reduced
without affecting the higher multipole correlations of the angular power spectrum of the temperature anisotropies. In the Ellipsoidal
Universe the Friedmann-Robertson-Walker metric  Eq.~({\ref{1}) is replaced by:
\begin{equation}
\label{6}
d s^2 \; = \; - c^2 \; dt^2 \; + \;  a^2(t) \; \left ( \delta_{ij} \; + \; h_{ij} \right )  dx^i  \; dx^j \; 
\end{equation}
with
\begin{equation}
\label{7}
 h_{ij}(t) \; = \;  - \; e^2(t) \; n_i \; n_j 
\end{equation}
where $e(t)$ is the ellipticity, and the unit vector $\vec{n}$ determines the direction of the symmetry axis. Moreover, at variance with the standard
cosmological model, the Ellipsoidal Universe model is able to account for large-scale CMB polarisation without invoking reionization processes.
Indeed, in our previous papers~\cite{Cea:2010,Cea:2014,Cea:2020} we were able to fix the eccentricity at decoupling and the polar angles 
$\theta_n$, $\phi_n$ of the direction of the axis of symmetry $\vec{n}$ such that the quadrupole temperature-temperature correlation matched
exactly the Planck 2018 value. We found~\cite{Cea:2020}:
\begin{equation}
\label{8}
 e_{dec}  \;   =  \;  8.32 \; \pm \; 1.32 \; \; 10^{-3}    \;   \; ,
\end{equation}
\begin{equation}
\label{9}
\theta_n  \;   \simeq  \;  73^\circ    \;  \;  , \; \;  \phi_n \;  \simeq   264^\circ  \;    \;   \; .
\end{equation}
We, also, showed that the quadrupole TE and EE correlations compared reasonably well to the Planck 2018 data. These results
allowed us to reach the conclusion that the Ellipsoidal Universe cosmological model not only were a viable alternative
to the standard cosmological model, but also it seemed to compare observations better than the $\Lambda$ Cold Dark Matter
cosmological model. In fact, recently we have found that the Ellipsoidal Universe cosmological model can account also
for the observed anomalous behaviour of the CMB two-point angular correlation function~\cite{Cea:2022}.
\section{The Hubble tension}
\label{s-3}
We address, now, the problem to see if the anisotropies in the universe spatial geometry can be able to alleviate the Hubble and
$S_8$ tensions. The Hubble constant can be inferred from the angular size of the sound horizon at recombination $\theta^*$ that,
in turns, is given by the ratio of the comoving sound horizon to the comoving angular diameter distance to the last-scattering surface:
\begin{equation}
\label{10}
 \theta^*  \; = \;  \frac{r_s(z^*)}{D_M(z^*)} \; \; ,    
\end{equation}
 $z^*$ being the redshift when the CMB radiation was last scattered. The comoving linear size of the sound horizon and the angular
 distance are linked to the expansion history of the Universe through:
\begin{equation}
\label{11}
r_s(z)  \; = \;  \int_z^{\infty}  \frac{c_s(z')}{H(z')}  \; dz' \; \;  
\end{equation}
and
\begin{equation}
\label{12}
D_M(z)  \; = \;  \int_0^z  \frac{c}{H(z')}  \; dz' \; \;  
\end{equation}
with $c_s(z)$ the speed of sound and $H(z)$ the Hubble constant at redshift $z$.
At early times, relevant for computing the sound horizon at recombination, in the $\Lambda$CDM model one can write:
\begin{equation}
\label{13}
H(z)  \;   \simeq  \;   H_0 \, \sqrt{\Omega_{\Lambda} \;  + \; \Omega_m \, (1 \, + \,  z)^3}  \; \; ,
\end{equation}
where    $\Omega_m$  and  $\Omega_{\Lambda}$  are the fractional densities of matter and dark energy satisfying the constraint:
\begin{equation}
\label{14}
\Omega_{\Lambda}  \;  + \; \Omega_m  \;  = \; 1   \; \; \; .
\end{equation}
So that we can rewrite Eq.~(\ref{10}) as:
\begin{equation}
\label{15}
\frac{c}{H_0} \,  \int_0^{z^*}  \frac{dz}{\sqrt{\Omega_{\Lambda} \;  + \Omega_m \, (1 \, + \,  z)^3} }  \; =  \; \frac{r_s(z^*)}{ \theta^*} \; \; . 
\end{equation}
Using the following values taken from Table~2 in Ref.~\cite{Aghanim:2021}:
\begin{equation}
\label{16}
\Omega_{\Lambda}  \; \simeq \;  0.689 \; \; , \; \;  \Omega_m \; \simeq \; 0.311
\end{equation}
and
\begin{equation}
\label{17}
z^* \; \simeq \; 1090 \; \; , \; \;  \theta^*   \; \simeq \;  1.041 \, \times \, 10^{-2}  \; \;  , \; \;  r_s(z^*) \; \simeq \; 144.6 \; Mpc \; \; ,
\end{equation}
we readily obtain from Eq.~(\ref{15}):
\begin{equation}
\label{18}
H_0  \; \simeq \; 67.9  \;  \; Km \, s^{-1} \, Mpc^{-1} 
\end{equation}
that agrees with the best-fitted Hubble constant Eq.~(\ref{2}).
Now, let us focus on the Ellipsoidal Universe model. In this case H(z) becomes:
\begin{equation}
\label{19}
H(z)  \;   \simeq  \;   H_0 \, \sqrt{\Omega_{\Lambda} \;  + \;  \Omega_m \, (1 \, + \,  z)^3   \; + \; \Omega_a(z) }  \; \; 
\end{equation}
since the source of anisotropy adds the term $ \Omega_a(z)$ due to a generic and unspecified anisotropic component related
to the cosmic shear~\cite{Campanelli:2011a,Campanelli:2011b}. Obviously, we have the constraint:
\begin{equation}
\label{20}
\Omega_{\Lambda}  \;  + \; \Omega_m  \; + \; \Omega_a(0) \; =  \; 1   \; \; .
\end{equation}
It turned out~\cite{Campanelli:2011a,Campanelli:2011b} that the cosmic shear is always smaller than unity. Moreover, the actual fraction
of energy associated to the anisotropic component is negligible with respect to those of matter and dark energy. Therefore, 
to a good approximation we will assume in what follows:
\begin{equation}
\label{21}
 \Omega_a(z) \;  \simeq  \; 0   \; \; .
\end{equation}
Nevertheless, it is worthwhile to mention the recent study presented in Ref.~\cite{Akarsu:2019}. The authors of Ref.~\cite{Akarsu:2019}
considered an anisotropic generalisation of the base $\Lambda$CDM model where the cosmic shear was assumed to
behave like a stiff fluid. Interestingly enough, they found that even with a tiny source of anisotropy the mean value of $H_0$ and
$\Omega_m$ are systematically larger than those in the case of the standard $\Lambda$CDM model, though with a rather low statistical
significance. Thus, we see that the results of Ref.~\cite{Akarsu:2019} are a first indication that a small anisotropy in the universe expansion rate 
tends to alleviate the $H_0$ tension.
In addition, we need also to take into account the cosmological aberration that affect the measurement of the angular size of the
sound horizon. To see this, we consider the null geodesic in the Ellipsoidal  Universe:
\begin{equation}
\label{22}
 c^2 \; dt^2 \; = \;  \;  a^2(t) \; \left ( \delta_{ij} \; - \; e^2(t) \, n_i \, n_j  \right )  dx^i  \; dx^j \; \; .
\end{equation}
First, we rewrite Eq.~(\ref{22}) as:
\begin{equation}
\label{22-a}
 c^2 \; dt^2 \; = \;  \;  a^2(t) \; \left ( 1 \; - \; e^2(t) \, (\vec{n} \cdot \hat{k})  \,  (\vec{n} \cdot \hat{k}') \right )  d \vec{x}^2  \; ,
\end{equation}
where $\hat{k}(\theta,\phi)$ and  $\hat{k}'(\theta',\phi')$ are the photon directions. We may consider $\theta \simeq \theta'$,
$\phi \simeq \phi'$, so that the average over the photon directions gives:
\begin{eqnarray}
\label{22-b}
 <  (\vec{n} \cdot \hat{k})  \,  (\vec{n} \cdot \hat{k}')  >  \;  \simeq \; \frac{1}{4 \pi} \;   \int_0^{2 \pi} d \phi    \int_0^{\pi} d \theta \sin \theta 
\left [ \sin \phi_n \sin \phi \cos (\theta_n - \theta) + \cos \theta_n \cos \theta \right ]^2  \; 
\nonumber \\
\; = \;  \frac{1}{12} \; \left [ 4  \cos^2 \theta_n \; - \; (\cos 2 \theta_n - 3) \sin^2 \phi_n  \right ] \; \;.  \hspace{4 cm} 
\end{eqnarray}
From this last equation we infer:
\begin{equation}
\label{23}
 c^2 \; dt^2 \; = \;  \;  a^2(t) \; \left [ 1 \; - \;  \frac{1}{12} \; \left [ 4  \cos^2 \theta_n \; - \; (\cos 2 \theta_n - 3) \sin^2 \phi_n  \right ]  \; 
  e^2(t) \right ]  d \vec{x}^2  \; \; .
\end{equation}
So that the comoving angular distance becomes:
\begin{equation}
\label{24}
D_M^{El}(z)  \; = \;  D_M(z) \; + \;     \delta D_M(z)
\end{equation}
with
\begin{equation}
\label{25}
 \delta D_M(z)  \simeq  - \, \frac{1}{24} \, \left [ 4  \cos^2 \theta_n  - (\cos 2 \theta_n - 3) \sin^2 \phi_n  \right ]  \,
\frac{c}{H_0} \,  \int_0^z  \frac{e^2(z')}{ \sqrt{\Omega_{\Lambda} \;  + \;  \Omega_m \, (1 \, + \,  z')^3} }  \; dz' \; . 
\end{equation}
Now, let us suppose that $\theta^*$ is the comoving angular diameter of  the last scattering surface. Thus, we get:
\begin{equation}
\label{26}
D_M^{El}(z^*) \, \theta^*  \; = \;  r_s(z^*) \; \; .
\end{equation}
Using  Eq.~(\ref{24}) we rewrite Eq.~(\ref{26}) as:
\begin{equation}
\label{27}
D_M(z^*) \, \theta^* (1 \; - \; \delta)  \; = \;  r_s(z^*) \; \; ,
\end{equation}
where:
\begin{equation}
\label{28}
\delta \; = \;  - \; \frac{\delta D_M(z^*) }{  D_M(z^*)}   \; \; .
\end{equation}
Equation (\ref{27}) tells us that the measured  angular  diameter assuming an isotropic spatial metric is:
\begin{equation}
\label{29}
\theta^*_{meas} \;  \simeq \; \theta^* (1 \; - \; \delta)  \; \; .
\end{equation}
Therefore we can write:
\begin{equation}
\label{30}
D_M(z^*) \; \simeq \; \frac{ r_s(z^*)}{ \theta^*_{meas}   (1 \; + \; \delta)}     \; \; ,
\end{equation}
where  $\theta^*_{meas}$ is given by Eq.~(\ref{17}).
Explicating Eq. (30):
\begin{equation}
\label{30bis}
\frac{c}{H_0} \,  \int_0^{z^*}  \frac{dz}{\sqrt{\Omega_{\Lambda} \;  + \Omega_m \, (1 \, + \,  z)^3} } 
\; \simeq \; \frac{ r_s(z^*)}{ \theta^*_{meas}   (1 \; + \; \delta)}     \; \; ,
\end{equation}
one sees that the parameter $\delta$ modifies the value of $H_0$. Moreover, combining   Eqs.~(\ref{25}) and (\ref{28})
 it is easy to check that $\delta  >  0$, so that Eq.~(\ref{30bis}) results  in an estimate of the Hubble constant $H_0$ 
 greater than that of the standard  $\Lambda$ CDM cosmological model.
 \\
\begin{figure}[t]
\centering
\includegraphics[width=0.90\textwidth,clip]{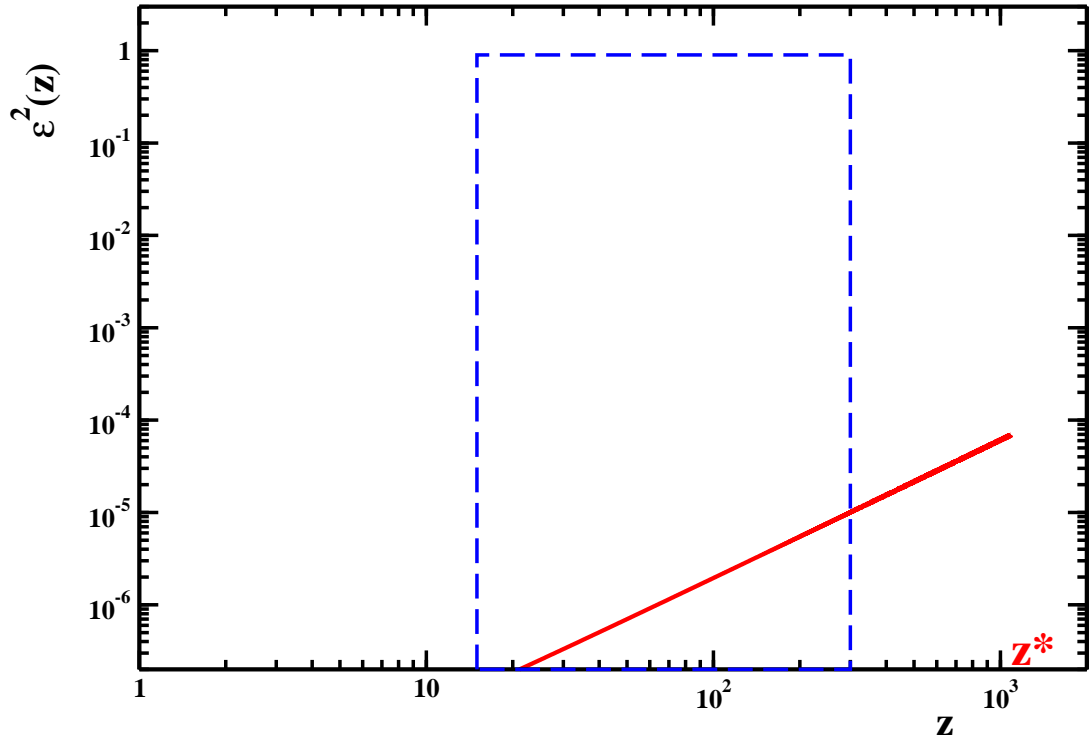}
\caption{\label{Fig1}  The (red) continues line is the ellipticity as a function of the redshift z 
 for the Ellipsoidal Universe model as given by  Eq.~(\ref{31}).  The (blue) dashed line
 is $\varepsilon^2(z)$ for the assumed  period of sizeable anisotropies during the  Dark Age, Eq.~(\ref{33}).
}
\end{figure}
In Ref.~\cite{Campanelli:2007} it was shown that an uniform magnetic field, a domain 
wall or a cosmic string gave rise to an anisotropic contribution to energy-momentum tensor
that, in turns,  were able to induce an anisotropic metric with planar symmetry. 
Moreover, it turned out that in any case $e^2(t) \sim a(t)^{3/2}$~\footnote{See Appendix A in Ref.~\cite{Cea:2014}}. 
Therefore, in the  matter-dominated era near decoupling we can write (see Fig.~\ref{Fig1}):
\begin{equation}
\label{31}
e^2(z) \; \simeq \; e^2_{dec} \;  \left [ \frac{1 + z }{ 1 + z^*} \right ]^{\frac{3}{2}}    \; \; , \; \; z \; \leq \; z^* \; \; .
\end{equation}
We obtain, then:
\begin{equation}
\label{32}
\delta \; \simeq \;  2.1 \; \times \; 10^{-7} \; 
\end{equation}
that, obviously, is too small to account for the $H_0$ tension. It is noteworthy that we may resolve the Hubble tension if we assume a 
finite period of sizeable anisotropies during the Dark Age, namely the period of time between the last scattering of the CMB radiation by 
the almost homogeneous plasma and the formation of the first star, Indeed, if we assume (see Fig.~\ref{Fig1}):
\begin{equation}
\label{33}
e^2(z) \; \simeq \; 0.90  \;  \; , \; \;   z_1 \; \simeq \; 15 \; \leq  \;  z \;  \leq \;   z_2 \; \simeq \; 300  \; \; ,
\end{equation}
we obtain:
\begin{equation}
\label{34}
\delta \; \simeq \;  3.4 \; \times \; 10^{-2} \; .
\end{equation}
After using Eq.~(\ref{30}), one gets:
\begin{equation}
\label{35}
H_0  \; \simeq \; 70.3  \;  \; Km \, s^{-1} \, Mpc^{-1}  \; \; ,
\end{equation}
that agrees with Eq.~(\ref{3}) within about two standard deviations. Note that the cosmic shear generated during such an extended 
period of time does not give rise to additional temperature anisotropies since the integrated Sachs-Wolfe effect vanishes:
\begin{equation}
\label{36}
\frac{\delta T}{T} \; \simeq \; - \; \frac{1}{2}  \int_{t^*}^{t_0}  dt \; \frac{\partial h_{ij}(t)}{\partial t} \; n^i \; n^j \; =  \; 0 
\end{equation}
where $t_0$ is the present time (z=0). Therefore, presumably, the only effects left should be some anisotropies in the matter distribution.
In this respect, it is interesting to note that recent studies (see  Refs.~\cite{Migkas:2020,Hu:2020,Krishnan:2021,Luongo:2021} 
and references therein) based on observations of quasars, supernovae and gamma ray bursts provided some evidences
for an anisotropic departure from the Friedmann-Robertson-Walker metric.
However, it is difficult to image physical  processes able to generate sizeable anisotropies in early times.
 One might think of cosmic defects drawn across space that crossed
through the Universe in the Dark Ages. Even thought such a mundane possibility is logically admissible, it should be evident that the
Ellipsoidal Universe cosmological model can accomodate values of $H_0$ larger than in the standard $\Lambda$CDM model, whilst not 
degrading the fits to the CMB data. After all the Ellipsoidal Universe model amounts to a simple anisotropic correction to the standard
cosmological model by replacing the spatially flat metric with the plane-symmetric Bianchi type-I metric. In this way one introduces
additional cosmological parameters that should be best-fitted to the precise Planck measurements.%
\section{The $S_8$ tension}
\label{s-4}
 Our previous discussion illustrated
how a tiny variation of the $\Lambda$CDM parameters resulted in an appreciable relaxation of the Hubble tension. This last point
can be better appreciated by looking at the $S_8$ tension. \\
As we said before, the $S_8$ tension arises from measurements with weak-lensing Planck data and redshift surveys. We, also, noticed
that the tension is mainly driven by $\sigma_8$. On the other hand, the amplitude of density perturbation $\sigma_8$ is tightly 
related to the primordial comoving curvature power spectrum amplitude $A_s$ defined, conventionally, at the pivot scale
$k_{pivot} = 0.05 \, Mpc^{-1}$. The CMB lensing reconstruction power spectrum constrains the late-time fluctuation amplitude
more directly in combination with matter density. Therefore, the dependence of the lensing power spectrum on $A_s$ can be eliminated
in favour of $\sigma_8$. The parameter dependence is given by~\cite{Ade:2016}:
\begin{equation}
\label{37}
\sigma_8^2  \; \propto \; A_s \; \Omega_m^{1.5} \; h^{3.5}  \; \; ,
\end{equation}
where $H_0 \, = \, 100 \times h \;  Km \, s^{-1} \, Mpc^{-1}$. \\
The observed CMB power spectrum amplitude scales with the primordial comoving curvature spectrum $A_s$. Actually, the
observed amplitude scales with $A_s \, \exp(-2 \tau)$ ($\tau$ being the optical depth) due to the scatterings of free electrons that
are present after reionization. Therefore, it is the combination  $A_s \, \exp(-2 \tau)$ that is well measured~\cite{Aghanim:2021}:
\begin{equation}
\label{38}
A_s \; \exp(-2 \tau) \; = \; 1.881 \; \pm \; 0.010 \; \times \; 10^{-9}   \; \; .
\end{equation}
In the standard $\Lambda$CDM cosmological model it is assumed that the large-scale CMB polarisation is due to reionization processes.
Thus, low-$\ell$ E-mode polarisation powers are dominantly produced by Thompson scattering of CMB photons off the free electrons
which are produced by reionization. So that the optical depth and the reionization redshift $z_{re}$ are well constrained by the large-scale
polarisation measurements~\cite{Aghanim:2021}:
\begin{equation}
\label{39}
 z_{re} \; = \; 7.82  \; \pm \; 0.71 \; \; , \; \;  \tau \; = \; 0.0561 \; \pm \; 0.0071   \; \; .
\end{equation}
Combining Eqs.~(\ref{38}) and (\ref{39}) one gets:
\begin{equation}
\label{40}
A_s \; = \; 2.105 \; \pm \; 0.030 \; \times \; 10^{-9}   \; \; .
\end{equation}
In the Ellipsoidal Universe model we showed~\cite{Cea:2010,Cea:2014,Cea:2020} that there is sizeable large-scale polarisation
signal without invoking reionization processes. Moreover, we found~\cite{Cea:2020} the CMB quadrupole TE and EE correlations were
in agreements with the Planck 2018 data. As a consequence, in the Ellipsoidal Universe cosmological model the optical depth
is not constrained, but it must be much smaller than the best-fit value Eq.~(\ref{39}). 
Therefore, in the Ellipsoidal Universe model we are led to the conclusion that the optical depth at reionizazion must be vanishingly small,
 i.e. $\tau \approx 0$. Thus, from Eq.~(\ref{38}) we obtain: 
\begin{equation}
\label{41}
A_s^{El} \;  \simeq  \; 1.881 \; \times \; 10^{-9}   \; \; ,
\end{equation}
that is smaller with respect to the standard cosmological model, Eq.~(\ref{40}). This, in turns, reduces the amplitude of density perturbation via
Eq.~(\ref{37})  and leads to:
\begin{equation}
\label{42}
S_8^{El} \;  \simeq  \; 0.780 
\end{equation}
that seems  to be close enough to Eq.~(\ref{4}) so as to eliminate the $S_8$ tension. \\
\section{Conclusions}
\label{s-5}
The Ellipsoidal Universe cosmological model amounts to a small anisotropic correction to the base $\Lambda$CDM cosmological model, that
was proposed several years ago~\cite{Campanelli:2006,Campanelli:2007} to explain the CMB quadrupole anomaly. Since then, we have
shown~\cite{Cea:2010,Cea:2014,Cea:2020}  that the Ellipsoidal Universe proposal can also  produce large-scale CMB E-mode
correlations in agreement with the latest Planck data. In the present note we are suggesting that the Ellipsoidal Universe cosmological
model should alleviate both the Hubble and $S_8$ tensions.  \\
\noindent
A planar Bianchi I model is a minimal deviation from the full set of  symmetries of the standard $\Lambda$CDM cosmological model. 
So fitting the observations with such a model does not introduce an insane number of  extra parameters and allows to test the cosmological principle. 
As such, the results of the present and previous papers indicated that the Ellipsoidal  Universe model looks attractive and promising. 
On the other hand, it should be clear that  it would be necessary to rerun all other cosmological tests and  especially to also check if the 
Ellipsoidal Model model allows us also to fit the angular power spectra. This becomes quite tough if one works with a non-isotropic 
 background since one would have  to  redo all the Einstein-Boltzmann hierarchy in Bianchi I  cosmological models.
To the best of our knowledge, in the literature there are no attempts to do so, but it is sure that it is necessary a lot of work. 
In fact, presently we are  considering the full set of Boltzmann equations in the Ellipsoidal Universe model and  try to solve, at least approximatively,
that equations.  We must admit that this is an hard task that require time. Nevertheless, this must be done if  one wish to compare quantitatively with the 
whole  CMB anisotropy data. Remarkably, up to now we were able to complete the calculations of the Boltzmann equations for
the photon distribution function. We have found that the full set of Boltzmann equations are confirming our previous results 
obtained by solving that equation at large scales. In particular,  the Boltzmann equations for the
photon distribution functions by taking into account the effects of the inflation produced primordial scalar perturbations and the anisotropy of the geometry  still satisfies:
\begin{equation}
\label{43}
\Delta T   \;  \simeq  \;  \Delta T^I   \;  + \; \Delta T^A   \;\;  
\end{equation}
where $\Delta T^I$ and  $\Delta T^A$ were the temperature fluctuations induced by the cosmological scalar perturbations and
by the spatial  anisotropy of the metric, respectively.  We hope to report further progress on this matter in future works.

\vspace{0.5cm}
\noindent
The Author declares that there is no conflict of interest. 

\vspace{0.5cm}
\noindent
 Data Availability Statement \\
\noindent
Data sharing not applicable to this article as no datasets were generated or analysed during the current study.
\vspace{0.5cm}
\noindent

\end{document}